\documentclass{llncs}
\usepackage{amsmath}
\usepackage{graphicx}
\usepackage{caption}
\usepackage{cite}
\usepackage{hyperref}
\usepackage{times}
\begin{document}
\title{Robotics and Integrated Formal Methods: \\ Necessity meets
  Opportunity\thanks{Work supported through EPSRC Hubs for Robotics and AI
    in Hazardous Environments: EP/R026092 (FAIR-SPACE), EP/R026173 (ORCA), and EP/R026084
    (RAIN).}}

\author{Marie Farrell \and Matt Luckcuck \and Michael Fisher}
\institute{Department of Computer Science, University of Liverpool, UK \\ \email{ \{marie.farrell, m.luckcuck, mfisher\}@liverpool.ac.uk }
}
\maketitle 
\begin{abstract}
Robotic systems are multi-dimensional entities, 
    combining both hardware and software, that are heavily 
    dependent on, and influenced by, interactions with the 
    real world. They can be variously categorised as embedded, 
    cyber-physical, real-time, hybrid, adaptive and even 
    autonomous systems, with a typical robotic system being
    likely to contain all of these aspects. The techniques for 
    developing and verifying each of these system varieties are often 
    quite distinct. This, together with the sheer complexity 
    of robotic systems, leads us to argue that diverse formal 
    techniques must be integrated in order to develop, verify, 
    and provide certification evidence for, robotic systems. 
    Furthermore, we propose the fast evolving field of robotics
    as an ideal catalyst for the advancement of integrated 
    formal methods research, helping to drive the field in new 
    and exciting directions and shedding light on the development 
    of large-scale, dynamic, complex systems.
\end{abstract}
\medskip

\section{Introduction}
Formal methods are used in a variety of domains to establish the
correctness of both hardware and software systems. Integrating formal
methods so that they may be used in a complementary fashion continues
to be a difficult challenge that is only exacerbated by the plethora
of languages, logics, theorem provers, and model-checkers available. In
this paper, we propose robotic systems as an ideal candidate for the
large-scale application of integrated formal methods. In fact, it is a
fast evolving area where only \textit{integrated} formal methods will
suffice. Further, the application of integrated formal methods in the
robotics domain will enhance integrated formal methods research and
promote their adoption for other large-scale, engineered systems.

Robotic systems are complex and multi-dimensional, with a wide range
of concerns: software, hardware, human control, autonomous agent
control, reconfigurability, etc. They present numerous challenges for
formal verification such as modelling a dynamic environment, providing
sufficient evidence for certification and trust, modelling multi-robot systems and, ensuring that
autonomous robots are safely reconfigurable and their decisions do not have dangerous side-effects. We discuss each of these challenges in
\S\ref{sec:challenges} as well as the current approaches to tackling
them. Our position is that the use of \textit{integrated} formal
methods can mediate these difficulties. In \S\ref{sec:ifmForRobotics},
we illustrate the benefits of integrated formal methods with respect
to these challenges and outline potential future directions for this
research. Finally, \S\ref{sec:conclude} provides concluding remarks.
The work cited here is not a complete list, it is drawn from a larger survey of formal specification and verification approaches for autonomous robotic systems \footnote{\url{http://tiny.cc/Luckcuck2018}}, which is still in progress.

\medskip

\section{Formal Approaches to Robotic Challenges}
\label{sec:challenges}
%
%

\noindent This section discusses some of the most crucial challenges to the
formal verification of robotic systems and how current formal
techniques approach them.  First, we discuss the verification of a
robotic system's interaction with an unknown and dynamic environment
(\S\ref{sec:environment}).  A further challenge is ensuring that
verification methods can provide suitable evidence either for
certification or to gain public trust
(\S\ref{sec:trustAndCertEvidence}). Certain types of robotic systems
present specific challenges and we discuss the challenges posed by
modelling multi-robot systems in
\S\ref{sec:swarms}. Finally, \S\ref{sec:autonomyAndAdaptation}
describes the challenges when verifying an autonomous or
reconfigurable robotic system.

Other challenges include the formal refinement of robotic system
specifications to implementable code and ensuring that this final
implementation corresponds to its specification in a provably correct
way. The heterogeneous nature of robotic systems where various
programming languages are used in the implementation of distinct
components of the system means that this final specification-to-code
step is not trivial. Since humans interact with these systems, it is
also necessary to model the human component of the system, however,
the verification of human behaviour is largely beyond the reach of current
formal verification approaches~\cite{IEEE:jnl,IJRR}.
 
\subsection{Modelling the Physical Environment}
\label{sec:environment}
To ensure that a robotic system can cope in real-world scenarios, it
must be able to react appropriately to an unknown and dynamic
environment. When formally modelling robotic systems, the environment
is often ignored~\cite{RAS:jnl} or assumed to be static and known,
prior to the robot's deployment~\cite{Moarref2017, IEEE:jnl,
  SCP:platoon}, which is often neither possible nor feasible in the
real world. Other approaches abstract away from the environment and
rely on predicates representing sensor data about the
environment~\cite{fisher_verifying_2013}. This insulates the
high-level control from the detailed environment, but leaves issues
such as sensor and actuator correctness to be dealt with.
Formal models of a robotic system's environment must bridge the
\textit{reality gap}, the difficulty of transferring models of the
environment to the real world~\cite{fisher_verifying_2013}. This is
especially problematic when real-world interactions can impact
safety. Reducing the impact of the reality gap often produces
intractable models which makes the verification task particularly difficult~\cite{Desai2017}.

Two popular approaches are to either model or monitor the physical environment. Temporal logics have been used to model robotic systems' environments. 
For example, safety rules and the environment of a robotic assistant captured in Probabilistic Temporal Logic (PTL)~\cite{IEEE:jnl}. 
However, as the rules and environment become more complex this approach may not be feasible due to the current limitations of model-checking techniques, unless the properties to be verified can be simplified. 

Specifying a monitor to restrict the robotic system to safe behaviours
within its environment reduces the verification burden, as only the
monitor needs to be verified~\cite{Machin2015}. For example, a robot's
environment can be captured by timed automata and safety properties
written in temporal logic~\cite{aniculaesei2016towards}. This can be
used to build a run-time monitor for the safety properties. This
combination of verification methods can help to handle the dynamic
environment.

Navigating an unknown and dynamic environment is a challenging task for robotic systems and a number of navigation
algorithms exist. However, not all can be employed in safety-critical
scenarios as they have not been verified
\cite{phan2015collision}. This suggests that there are limitations of current formal methods to verify these algorithms, and also
leads to hybrid approaches that have high computational complexity.
KeYmaera is a hybrid theorem prover that has been used to verify both the discrete and
continuous behaviour of robotic vehicle navigation using
differential dynamic logic for hybrid systems
\cite{mitsch2013provably}.

\subsection{Trust and Certification Evidence}
\label{sec:trustAndCertEvidence}
Robotic systems are often safety-critical, such as those in nuclear
or aerospace applications, and so require certification. Other
robotic systems operate in unregulated areas that require public
trust, such as domestic assistants. Emerging robotic systems, like
autonomous vehicles, require both certification and public
trust. Therefore, ensuring that formal verification of robotic systems
can provide appropriate trust and certification evidence is crucial. Generally, robotic
systems development provides insufficient evidence for certification
and public trust, which can hamper their
adoption~\cite{IEEE:jnl}. This is an area where integrating formal
methods with current non-formal engineering techniques may prove
fruitful.

Further to extensive testing, safety cases are generally used to
provide evidence for certification bodies. A safety case is a
structured argument that is supported by a collection of evidence
providing a compelling, comprehensible and valid case that a system is
safe for a given application in a specific
environment~\cite{denney2014automating}. Automating the generation of
such documentation is a challenging task that must account for
heterogeneous content such as physical formulae from the design of the
physical system, maintenance procedures, and software (which itself,
may be of a heterogeneous nature). Recent work in this area includes a
methodology for automatically generating such safety cases for the
Swift pilotless aircraft system using a domain theory and AUTOCERT
\cite{denney2014automating}.

Formal methods can provide suitable evidence for certification. For
example, Isabelle/HOL and temporal logic have been used to formalise a
subset of traffic rules for vehicle
overtaking~\cite{rizaldi2017formalising}. Furthermore, a
model-checking approach has been used to capture the rules and
expectations of pilots in order to to provide certification evidence
for an autonomous pilotless aircraft \cite{JAIS:jnl}. Here, it is
verified that the agent controlling the aircraft (in place of a pilot)
preserves the rules and recommendations specified by the Civil Aviation Authority (captured as temporal logic
formulae).

There are currently no guidelines to help developers choose the most suitable formal method to verify their
system~\cite{kossak2016select}. Similarly, regulators and
certification bodies are often hesitant to suggest suitable formal
methods for safety-critical systems -- though guidance has started to
appear more recently. Regulators, developers, and academia thus face
the challenge of how to determine suitable and robust formal methods
for particular types of robotic system.

\subsection{Multi-Robot Systems}
\label{sec:swarms}

Historically, the development of multi-robot systems has taken inspiration from biological systems such as swarms of insects.  Robot swarms are difficult to develop because they are programmed at the microscopic level (that of individual robots) but are intended to exhibit emergent, macroscopic behaviour (at the level of the whole swarm). They are often developed bottom-up, using trial and error to form a swarm with the desired emergent behaviour~\cite{Moarref2017}. Ensuring that macroscopic behavioural requirements (or restrictions) are implemented (or obeyed) at microscopic level can be difficult because of their different abstraction levels.

Robot swarms can be quite large, and so a challenge when verifying robot swarms using current model-checking techniques is state space explosion caused by the large number of concurrent, interacting agents and the system's dynamic environment~\cite{Akhtar2014}. This can be mitigated by making use of the homogeneity of the swarm's robots, for example by exploiting symmetry reduction~\cite{Antuna2015} or abstracting the swarm to a single state machine with a counting abstraction~\cite{RAS:jnl}. However, both of these approaches only consider swarms with homogeneous behaviour. 

The emergent behaviour of robot swarms can be captured using temporal logic~\cite{winfield_formal_2005} and often lend themselves to probabilistic models. In particular, PRISM has been used to encode probabilistic state machines which can then be checked for properties specified in a probabilistic temporal logic~\cite{RAS:jnl}. 

Further, robotic systems may consist of a team of heterogeneous robots. In hazardous environments such as nuclear plants, it is conceivable that multiple robots may be required in order to complete a specific task. Each robot would have a distinct role, for example, a robotic arm to examine a piece of debris and a more mobile robot to monitor, calibrate and mend the robotic arm should it malfunction. The robot team can be verified at the macroscopic level, but at the microscopic level each robot must be verified individually.  For example, one robot in the team may be characterised by a verifiably  correct Z specification, whereas another may be verified by model-checking its source code. The behaviour of the robot team might be different to the behaviour of each individual robot and thus another approach may be used to verify the team's behaviour.  It is not clear how best to link the verification approaches taken at these different levels of abstraction because the approaches most amenable to the verification of each individual robot might be different. 

\subsection{Adaptation, Reconfigurability and Autonomy}
\label{sec:autonomyAndAdaptation}

A self-adaptive system continually alters its behaviour in a feedback loop that is driven by its environment. A literature survey found that there are no standard tools for the formal modelling and verification of self-adaptive systems~\cite{Weyns2012}. Of the tools surveyed, 30\% use model-checking. One avenue of research suggests using (both semi-formal and formal) models to check run-time behaviour~\cite{Cheng2014}. This agenda considers approaches such as automatic test case generation and formal model-checking. The aim being to reduce state explosion by quantifying as many variables as possible at run-time.

Related to this is the notion of a reconfigurable system, which senses its environment and makes a \textit{decision} about how best to reconfigure itself to suit changes in its requirements or the environment. Reconfiguration is essential for ensuring the fault tolerance of safety-critical robotic systems~\cite{tarasyuk2012formal}. There are two key open questions when applying formal methods to these systems: (1) how to specify and analyse a configuration, and (2) how to compare different configurations of the same system~\cite{Morse2016}? 
The design of reconfigurable hardware has received much attention, but autonomous software reconfiguration remains a challenge~\cite{bi2008development}. One approach involves building a flexible control system that can reconfigure itself once a fault is detected~\cite{braman2007safety}. Z models can be used to describe an arbitrary reconfigurable system \cite{Weyns2010}. The model provides a method for describing and comparing different configurations of the system's architecture.

Since reconfigurability requires the system to make an autonomous
decision as to how best to reconfigure itself, it is vitally important
that the decisions made by the system are \textit{rational}, meaning
that the system can explain its reasoning. This leads us to model the
motivations and decisions of the system, ideally as first class
objects \cite{fisher_verifying_2013}. 

Agent-based systems are one way of describing autonomy; there are many different models of agent systems, based on different models of autonomy. Agents are used to model a robot's interactions with other actors, its environment, and the physical environment itself. For example, probabilistic temporal
logics have been used for modelling an autonomous mine detector robot, controlled by an agent, and its environment~\cite{Izzo2016}. The model of the agent can be used for both design- and run-time verification.

A formal model of a style of agent system using Z has been devised that gives its agents a formal semantics \cite{d2004dmars}. The interactions of multiple interacting agents can also be modelled using finite state machines. These are converted into Alloy specifications for automatic verification~\cite{podorozhny2007verification}. The size of these specifications meant that keeping the models tractable was challenging.

Relating agent programs, written in an agent programming language, and agent (verification) logics remains an open problem. One approach has been to define an agent programming theory combining an agent programming language and verification logic~\cite{hindriks2009toward}. Program model-checking such as the Model-Checker for Multi-Agent Systems (MCMAS), has been used to verify heterogeneous agents interacting with an environment~\cite{choi2015verification}; and Agent Java PathFinder (AJPF)~\cite{Dennis2012}, which can model-check programs written in a particular style of agent language. 
\medskip

\section{Integrated Formal Approaches to Robotic Challenges}
\label{sec:ifmForRobotics}

\noindent 
In \S\ref{sec:challenges}, we outlined the challenges encountered when
developing reliable robotics and a number of current (non-integrated)
approaches to addressing them. It is clear that only by using a
combination of specialised tools and methodologies can we achieve a
high level of confidence in software. For example, the NASA Remote
Agent uses specialized languages for each of the planner, executive,
and fault diagnosis components \cite{simmons_towards_2000}.

There is currently no general framework integrating formal methods for robotic systems. However, \S\ref{sec:currentifm} describes recent trends and some notable bespoke examples of integrated formal methods (iFM) for robotics. In \S\ref{sec:future} we discuss how robotics and iFM can benefit from one another. We do not ignore the important role that validation techniques, such as testing and simulation, play in the development of robotic systems. These too, should be integrated into the development process to be used alongside formal methods~\cite{IJRR}.

\subsection{Adopting iFM for Robotics}
\label{sec:currentifm}

Current approaches to formal verification in robotics typically centre around one tool or technique that is suited to verifying properties of a particular type (concurrency, probability, etc.). It is clear from the increasingly complex nature of robotic systems that this is not a sustainable approach to ensuring the correctness of these systems. These approaches suffer from a number of drawbacks that are mostly caused by the limitations of their logic or the tool being used. 

A comparison of four different specification formalisms (CSP, WSCCS, Unity Logic, and X-Machines) for specifying and verifying emergent swarm behaviour, concluded that a blending of these formalisms offered the best approach to specify emergent swarm behaviour as none was sufficient in isolation \cite{Hinchey2005}. This claim is further supported by the use of MAZE (an extension of Object-Z for multi-agent systems) that uses Back's action refinement to facilitate a top-down development process of the swarm, from the macroscopic to microscopic level~\cite{smith2014maze}. 

It is therefore clear that the use of iFM can help to mediate the issues surrounding the development of robotic systems as has been illustrated by the following bespoke examples.  
We outlined the importance of reconfigurable systems in \S\ref{sec:autonomyAndAdaptation}, and a combination of Event-B and the PRISM model-checker has been used to derive a reconfigurable architecture for an on-board satellite system \cite{tarasyuk2012formal}. The combination of these formal notations allows not only for the formal specification and derivation (via refinement) of the system in Event-B, but also the probabilistic assessment of its reliability and performance using PRISM.

The combination of AJPF for agent verification, Uppaal for timing properties and, spatial reasoning has been used to verify the procedures for a driverless car joining and leaving a vehicle platoon \cite{SCP:platoon,kamalimodular}. This work verifies the cooperation between the vehicles, and the abstract behaviour of the real physical vehicle. Related work uses CSP$\|$B to correctly model a real physical platooning vehicle \cite{cspb}.

Finite State Processes (FSP) and $\pi$ADL ($\pi$-calculus combined with the Architecture Description Language) have been combined to capture safety and liveness properties of multi-agent robotic systems \cite{Akhtar2014}. The FSP specifications of the relevant safety and liveness properties are transformed into Labelled Transition Systems, then the agent programs and architecture (described in $\pi$ADL) are checked to see if they satisfy the required properties. Designed as a generic notation for modelling robotic systems, RoboChart integrates the process algebra  CSP with a graphical timed state machine notation~\cite{Ribeiro2017}. This allows graphical visualisation of the specification and automatic model-checking of its behaviour. The use of a process algebra in these cases is ideal for modelling communication across a channel with $\pi$ADL and timed state machines, respectively, providing a robust model of the system's state that could not be achieved using a process algebra in isolation.

Furthermore, several views of the same system or component often
require integration of analysis, even for one specification
element. For example, model-checking, model-based testing, and user
evaluation have all been applied to the same robotic system
\cite{IJRR}. Each, however, works at a very different level of
abstraction and formality and so the challenge here is to integrate
this breadth of techniques in a holistic framework.

\subsection{Future Directions for iFM}
\label{sec:future}

The benefits of iFM are well known. Specifically with respect to robotics, iFM can: (1) enable us to capture detailed physical environments by combining static and dynamic models; (2) provide a formal mechanism for linking the macroscopic and microscopic levels of multi-robot systems; (3) provide robust evidence for trust and certification, and; (4) express the complex properties of adaptive, reconfigurable, and autonomous systems. This unique set of challenges posed by robotics provides tangible targets for iFM researchers. Thus, robotics can benefit from and be a catalyst for iFM research, and the adoption of these techniques for large-scale, dynamic, and complex systems.

Integrating multiple approaches to verification for systems in the safety-critical domain presents its own set of familiar challenges, such as increased complexity and ensuring the correctness of the integrated model. Until now, these challenges have generally been addressed using theoretical frameworks, small case studies, and prototype tools. Robotic systems are a complex and practical field where iFM is crucial to provably correct advances and adoption. Further to these challenges, usability is a concern from both perspectives. Firstly, the iFM community is tasked with providing a set of robust tools that are intuitive and usable for the developers of robotic systems. Secondly, robotic systems should be developed with iFM in mind using a set of standardised, modular constructs that are amenable to iFM.

Combining formal methods with different strengths and weaknesses, such as exhaustive model-checking and proof-based methods, provides a useful balance of complexity and robustness. While model-checking exhaustively examines the system's state space to check if a property is preserved, formal proofs provide a step-by-step mathematical argument as to why the property holds. Although, both model-checking and theorem proving generally involve abstracting to a formal specification in order to verify the system, an advantage of model-checking is that it can be used directly on the implemented code, whereas theorem proving cannot. In contrast, theorem proving techniques do not suffer from the explosion in state space that limits the complexity of the properties that can be verified using model-checking. Formalisms that support formal refinement of specifications, such as Event-B, facilitate a verification process that provides a proof of the properties that are verified at each level of abstraction. Integrating model-checking and proof-based approaches to verification will provide fast identification of bugs and a list of the properties that are verified using model-checkers, as well as robust mathematical arguments for correctness in the form of proofs. These techniques, in combination, can thus provide the more robust certification evidence required for robotic systems.

Robotic systems are layered entities containing both hardware and software components. In general, each layer is built upon the lower layers and assumes that they behave correctly. In this scenario it is likely that the formalisms and tools used for both the verification and implementation of each layer are different. This presents a huge challenge for ensuring the correctness of the entire system, and in particular, verifying the interactions between these layers.
Contemporary robotics software is often highly modular, with components loosely connected. For example the Robot Operating System (ROS) allows architectures comprising heterogeneous components, written in different programming languages, which can interface with a range of hardware and software components~\cite{quigley2009ros}. There will undoubtedly be different techniques relevant to, and optimal for, the verification and validation of each of the different components. These include stochastic analysis of a learning component, a range of testing techniques for a vision component, or model-checking of an autonomous decision-maker. These must all be combined to provide a coherent and comprehensive analysis of the whole system.

We propose the verification of middleware architectures, such as ROS, as an ideal starting point for this research agenda. For example, the specification and verification of individual ROS nodes using pre- and post-conditions or assume-guarantee clauses, written using a heterogeneous collection of logics, would prove useful here. This  approach could also aid in the verification of heterogeneous teams of robots as discussed in \S\ref{sec:swarms}.
It is clear that a common framework for translating between, relating, or integrating different formal methods and validation techniques will prove useful. Such a framework would enable easy conversion between formalisms and verification tools. This would facilitate the use of heterogeneous models that are each suited to a particular type of behaviour or property. Moreover, the use of iFM can save time in the development process by avoiding duplicate specifications and exploiting different types of verification tools for proving different properties of the same system.

\section{Conclusions}
\label{sec:conclude}

\noindent Robotic systems are inherently multi-dimensional entities
that combine both hardware and software components that interact with
humans and the physical world. These systems can be modelled in a
variety of ways and thus must integrate verification and validation
techniques from the fields of embedded, cyber-physical, real-time,
hybrid adaptive and even autonomous systems. In
\S\ref{sec:challenges}, we discussed the challenges that are
encountered when developing certifiably correct robotic systems and
the current formal approaches to tackling them. It is clear that
current (non-integrated) formal methods are not robust enough,
particularly in isolation, to ensure the correctness of these
systems. Although not without its challenges, in
\S\ref{sec:ifmForRobotics}, we have illustrated the benefits of
employing \textit{integrated} formal methods in this
setting and outlined future directions for this work. Furthermore, we have argued that although robotics actually
necessitates the use of integrated formal methods, integrated formal
methods can utilise robotics as a viable and impactful means for
advancing integrated formal methods research and their adoption for
large-scale, complex systems. The challenges that we have outlined throughout this paper can be achieved through funding streams such as EPSRC at a national level and Horizon 2020 at an international level.

\newpage
\bibliographystyle{abbrv}
\bibliography{bibliography} 

\end{document}